\documentclass[aps,prl,reprint]{revtex4-1}
\usepackage{amsmath}
\usepackage{amssymb}
\usepackage{graphicx}
\usepackage{mathrsfs}
\usepackage{natbib,hyperref}
\usepackage{mathtools}

\renewcommand{\FL}{f_{\lambda}}
\newcommand{\FQ}{f_{q}}
\newcommand{\FP}{f_{p}}

\newcommand{\intsmile}{\mathrlap{\int}{\smile} \,\,}
\newcommand{\intfrown}{\mathrlap{\int}{\frown} \,\,}

\newcommand{\f}{\mathbf{f}}

\newcommand{\lpc}{\lambda_\mathrm{pc}}

\renewcommand{\Pr}{\mathrm{Pr}}

\renewcommand{\dfrac}[2]{\frac{\mathrm{d}#1}{\mathrm{d}#2}}

\begin{document}

\title[Open system control of dynamical transitions under the generalized KNH theorem]{Open system control of dynamical transitions under the generalized Kruskal-Neishtadt-Henrard theorem}

\author{D M Fieguth and J R Anglin}

\address{State reaserch center OPTIMAS and Fachbereich Physik, Technische Universit\"at Kaiserslautern,D-67663 Kaiserslautern, Germany}

\date{\today}

\newcommand{\p}{^\prime}
\renewcommand{\d}[1]{\!\!\mathrm{d}#1\;}
\renewcommand{\dfrac}[2]{\frac{\mathrm{d} #1}{\mathrm{d} #2}}
\newcommand{\Cvec}{\f}
\newcommand{\intd}{\int\!\mathrm{d}}
\newcommand{\pfrac}[2]{\frac{\partial #1}{\partial #2}}
\newcommand{\Bgam}{\mathbf{B}_\gamma}
\newcommand{\psl}{\mathcal{P}^{(\lambda)}}
\renewcommand{\div}{\operatorname{div}}
\newcommand{\xvec}{\mathbf{x}}

\begin{abstract}
Useful dynamical processes often begin through barrier-crossing dynamical transitions; engineering system dynamics in order to make such transitions reliably is therefore an important task for biological or artificial microscopic machinery. Here we first show by example that adding even a small amount of back-reaction to a control parameter, so that it responds to the system's evolution, can significantly increase the fraction of trajectories that cross a separatrix. We then explain how a post-adiabatic theorem due to Neishtadt can quantitatively describe this kind of enhancement without having to solve the equations of motion, allowing systematic understanding and design of a class of self-controlling dynamical systems.
\end{abstract}

\maketitle
\section{I. Introduction}
From satellite capture in astrophysics to chemo-mechanical reactions, dynamical transitions induced by slow parameter change are important phenomena in many physical settings. Adiabatic approximations that are usually highly accurate can break down around such barrier-crossing events, and seemingly small effects can make the difference between a transition that occurs rarely, for only a small set of initial conditions, and one that occurs more reliably.

Dynamical transitions of this kind are typically described within adiabatic theory, assuming that the Hamiltonian which governs the system has some slow, fixed time dependence, through an ``external'' parameter which slowly changes in a predetermined way  \cite{Dobbrott,Henrard,Neishtadt,Chirikov2,Timofeev, Cary,Hannay,Cary2,Elskens,Chernikov,Chow,Lu,Benisti,Benisti2,Chernikov2,Carioli,Cappi}. The adiabatic approximation breaks down near a separatrix, and so post-adiabatic methods are needed to describe whether and how the system may cross an adiabatic separatrix, which usually implies a qualitative change in its behavior, analogous to the change between rotation and libration in a physical pendulum, or to the initiation of a chemical reaction. If we study dynamical transitions for their own sake, rather than merely as failures of adiabatic theory, the slow ``sweep'' of the time-dependent parameter may indeed be considered as, or as if it were, a control strategy, of which the purpose is to induce the transition.

By applying Liouville's theorem as well as adiabatic analysis, the Kruskal-Neishtadt-Henrard theorem (KNH) derives the probability of a dynamical transition into a given region of phase space in terms of the rate at which the area of that region is changing (due to the slow change of the Hamiltonian). It has recently \cite{Liouvillepaper} been pointed out that this relation can be the basis of a class of ``blind'' control strategies: by engineering a slowly time-dependent Hamiltonian to increase the growth rate of a target region, the set of initial conditions which lead to transitions into the target region can be enlarged, without having to monitor or control fast degrees of freedom.

Realistically, a time-dependent ``external" parameter is usually a dynamical degree of freedom in its own right---possibly a collective coordinate of a large reservoir---which is simply slow and heavy enough for the back-reaction of the system upon it to be neglected, as in the Born-Oppenheimer approximation of molecular physics. Even if back-reaction is only a small perturbation, however, time-dependent parameters frequently do react at least slightly to the system's evolution. The slowly changing parameter can be said to record some information about the system, and modify its effect on the system accordingly, even if these reaction effects are only small. The evolution within the system phase space alone, without considering the evolution of the parameter, may thus be non-Hamiltonian. Beyond the zeroth-order Born-Oppenheimer approximation of a predetermined external parameter, therefore, slowly sweeping a parameter amounts to a strategy of \emph{open-system} control.

Our purpose in this paper is to point out that even small back-reaction effects can be important in controlling separatrix-crossing dynamical transitions. They may offer an opportunity to enlarge the range of initial conditions for which a desired transition occurs, potentially converting a rare fluke into a reliable consequence. 

We demonstrate this firstly in Section II with an example, comparing two models in which the actual time-dependences of a parameter remain very close to each other, but in one case back-reaction induces small correlations between system and parameter, which induce dynamical transitions that do not occur in the other model. We then show in Section III that this advantage of open-system control exists well beyond our illustrative example, by using Neishtadt's generalization of the KNH theorem (GKNH)\cite{Neishtadt2,Neishtadt3} to compute the enhancement of transition probabilities in generally similar scenarios. We conclude in Section IV with a brief discussion.

\section{II. Enhancing transition probability with open-system control: an example}

\subsection{II.1 Dragging as a control task}

Our example to show the general effect is a particle in one dimension, with phase space coordinates $q,p$ as usual, subject to a potential which depends on a time-dependent parameter $\lambda(t)$. In particular the potential is sinusoidal, with a $\lambda$-dependent amplitude that slowly ramps on and off. At the same time as the potential ramps on and off in strength, it also moves back and forth in space, in dependence on $\lambda$. The goal of this two-fold $\lambda$ dependence, as a control task, is to drag the particle back and forth over a certain distance, by capturing it into a bound orbit in one of the wells of the moving sinusoidal potential. 

Since the potential initially has vanishing strength, the particle is not bound initially; as the potential strength grows, a phase space region of bound orbits grows, but the particle is not initially in it. To succeed in the control task, the particle must cross a dynamical separatrix into this bound-orbit region. 

For some initial conditions it does this, while for others it does not. We compare the phase space areas of initial conditions that lead to capture and dragging for two different forms of $\lambda(t)$ time dependence. The two forms will differ only very slightly as functions of time, but they will differ qualitatively in precisely how their time dependences are determined. In one case, it is preordained, while in the other it depends weakly on the instantaneous momentum $p$ of the particle.

\subsection{II.2 Co-moving frame Hamiltonian}
As a simple case we will take the potential's instantaneous velocity to be the parameter $\lambda$, on which the potential's instantaneous strength also depends. To simplify our description of the problem we will work in the co-moving frame of the moving potential, where our $\lambda$-dependent Hamiltonian reads
\begin{equation}
    H(\lambda,q,p)=\frac{(p-\lambda)^2}{2}-\beta^2(\cos(q)+1),\label{eq:HAM}
\end{equation} with $\beta=\sqrt{g}\exp\left(-\frac{\alpha}{4}\lambda^2\right)$.
In these co-moving frame coordinates, success in the control task of dragging means keeping $q$ constant. Since dragging occurs when the particle is captured into a potential well, a successful dragging will also involve small oscillations of $p$ around $\lambda$ as $\lambda$ slowly changes. Failure to capture the particle and drag it will appear instead as $q$ changing with $\lambda$, as the particle fails to move with the potential, while $p$ remains constant. 

\begin{figure}
	\centering
	\includegraphics*[width=0.45\textwidth]{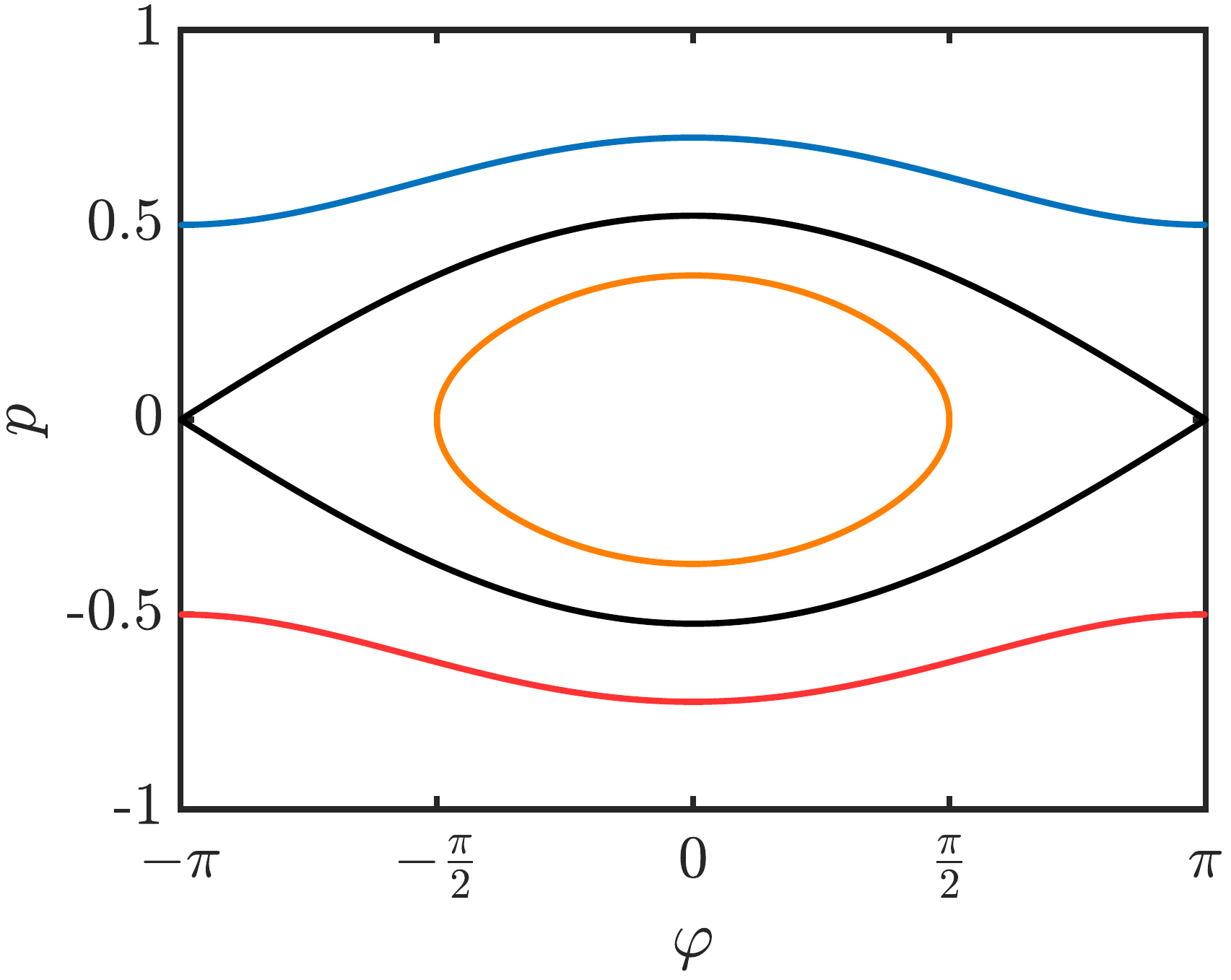}
	\caption{\label{fig:Cateye} Instantaneous energy contours in phase space for $\lambda=0$, when the separatrix is at its largest size. Shown are orbits with  energies $H(0,\pi/2,0)$ (orange) , $H(0,\pi,0)$ (the separatrix, blue) and $H(0,\pi,0.5)$ (red) for $g=(\pi/4)^2/9,\alpha=1$.}
\end{figure}

Captured and non-captured orbits in phase space can be recognized in the energy contours of $H$ for a fixed value of $\lambda$, as shown in Figure~\ref{fig:Cateye}.
We see the separatrix (black) and an orbit inside it (orange), as well as orbits below (red) and above (blue) the separatrix. The eye-shaped separatrix becomes vanishingly small for large $|\lambda|$, as the Gaussian factor the potential strength becomes small, but in general there is a separatrix at any $\lambda$, separating phase space into three regions. Since the corners of the separatrix are at the unstable fixed points $p=\lambda$, $q=\pm\pi$, the separatrix is a contour with zero energy and a parametrization is given by
\begin{equation}
p_\pm(q,\lambda)=\lambda\pm \beta\sqrt{2+2\cos q}\;,\label{eq:psep}
\end{equation}
enclosing the area
\begin{equation}
A_\mathrm{sep}(\lambda) = 16\beta\;.\label{Asep}
\end{equation}
Initially $\lambda$ will be negative, and it will steadily increase, so that the separatrix slowly grows and then shrinks, while simultaneously moving upward in $p$. 

\begin{figure*}
	\centering
	\includegraphics*[width=1\textwidth]{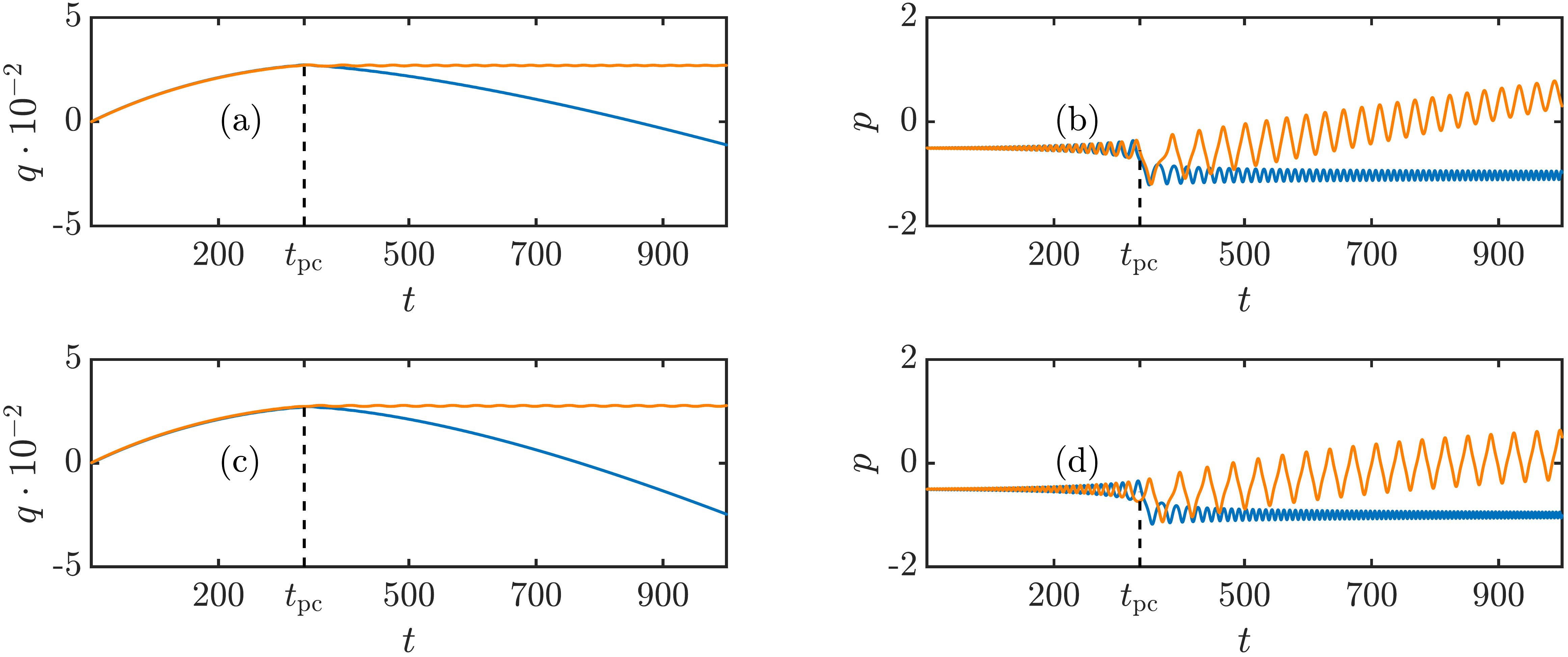}
	\caption{\label{outcomes} Time evolution of $q$ (panel (a)) and $p$ (panel (b)) for trajectories of the responsive case transitioning to different phases of motion. The same qualitative behaviour can be seen for the unresponsive case, $q$ (panel (c)) and $p$ (panel (d)). Slight differences between the orange curves in (b) and (d) can be seen around $t_\mathrm{pc}$, and also at late times in the decreasing amplitude of the orange-curve oscillations in (b), with no corresponding amplitude decrease in (d). The constants are $g=(\pi/4)^2/9,\alpha=1,\gamma=1$ and $\varepsilon=0.002$.}
\end{figure*}

The particle will begin outside the separatrix, above it. If the control task succeeds, the particle will enter the separatrix as the separatrix migrates upward, and orbit within it for some significant time, moving upward with it in $p$. If the task fails, the particle is never captured into the separatrix, but is left behind in the region below the separatrix after it passes. Figure~\ref{outcomes} shows examples of both kinds of trajectories, which differ only slightly in their initial conditions. Panels (a)-(b) and panels (c)-(d) are for two different evolutions of $\lambda$ we will discuss in the next section.

\subsection{II.3 Parameters with and without back-reaction}
As a simple example of parameter time dependence which responds to the system, our first case of $\lambda(t)$ (``the responsive case'') obeys
\begin{align}
    \dot \lambda=\varepsilon(1+\gamma(p-\lambda)).\label{eq:lamdotExact}
\end{align}
We will contrast the effect of this $\lambda(t)$ with that of the ``unresponsive case'' $\lambda\to\bar{\lambda}(t)$ for
\begin{equation}
    \bar{\lambda}=\left(R-S\right)\mathrm{e}^{-\varepsilon\gamma t}+S, \label{eq:TimeAvUp}
\end{equation}
with constants $R$ and $S$. Unlike $\lambda(t)$, $\bar{\lambda}(t)$ does not respond to the instantaneous $p(t)$ of the system, but merely follows a predetermined time evolution, relaxing slowly from initial $R$ to final $S$. The reason for this specific form of $\bar{\lambda}(t)$ is that it allows us to investigate the specific effect of the \emph{responsiveness} of $\lambda(t)$ on dynamical transitions, because by tuning $R$ and $S$ appropriately to the initial conditions $\lambda(0)$ and $p(0)$, we can make the predetermined evolution of $\bar{\lambda}(t)$ conform closely, though not exactly, to the actual evolution of $\lambda(t)$, right through the time within which the capturing transition either occurs, or does not.

\begin{figure*}
	\centering
	\includegraphics*[width=1\textwidth]{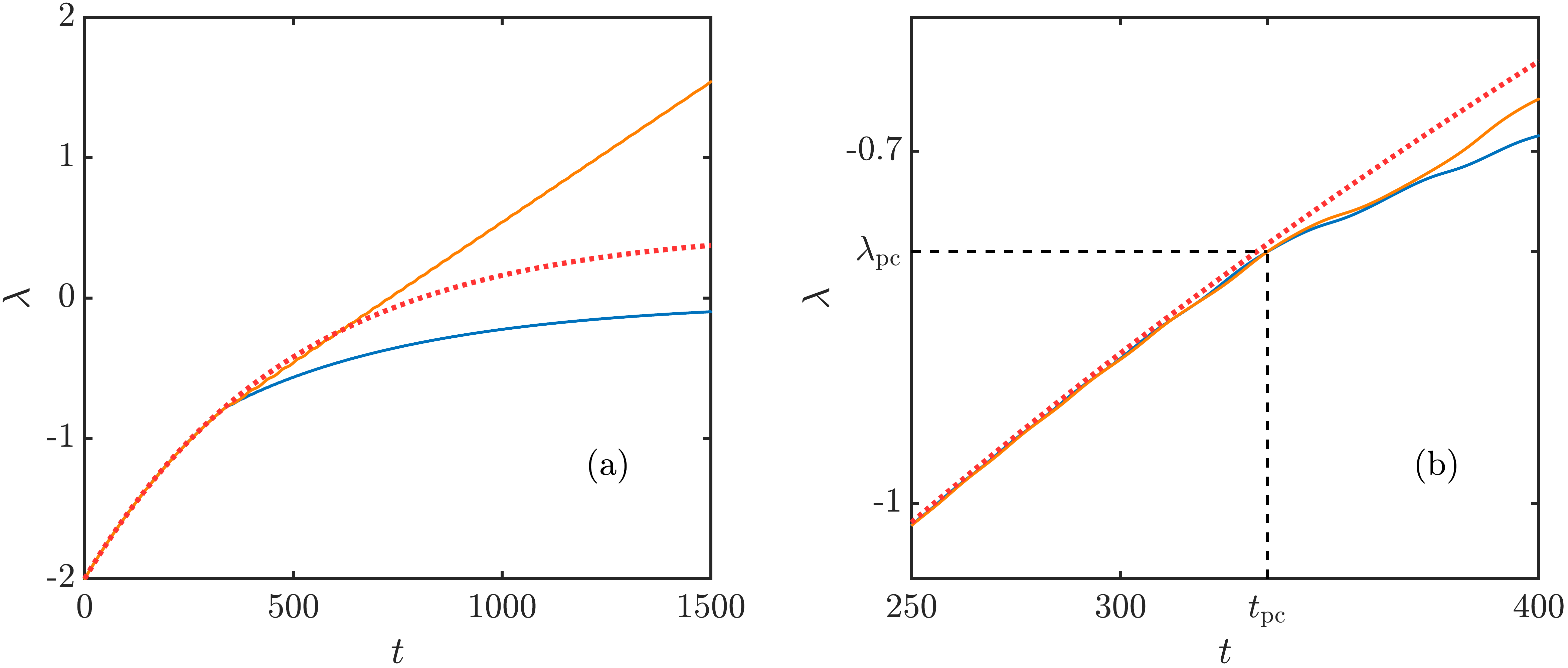}
	\caption{\label{fig:EVOLAM} Time evolution of $\lambda$ for the same initial conditions as in Figure~\ref{outcomes}. The orange trajectory enters the separatrix and the blue one does not. The red dotted line is approximation  \eqref{eq:TimeAvUp}. Panel (b) depicts the same trajectories but zoomed in around $\lpc$.}
\end{figure*}

The particular form of $\bar{\lambda}(t)$ allows us to do this, because until the system approaches the separatrix, ordinary adiabatic theory will accurately describe $q(t)$ and $p(t)$, and we can therefore also compute $\lambda(t)$ self-consistently, to high accuracy for small $\varepsilon$. As often with adiabatic evolution, the result is equivalent to averaging away small-amplitude, high-frequency components of $\lambda(t)$. This analysis is presented fully in Appendix B, but can be outlined here briefly. If the particle is not (yet) captured, then the only effect of the sinusoidal potential on $p(t)$ will be ``speed bump'' modulations which average away, leaving the constant average $\bar{p}(t) = p(0)$. Substituting $p(t)\to p(0)$ in (\ref{eq:lamdotExact}) then leads to the self-consistent solution $\lambda(t)\to\bar{\lambda}(t)$ for
\begin{equation}
   R=\lambda(0)\qquad S=p(0)+\frac{1}{\gamma}\;.\label{eq:RS}
\end{equation}
How well the resulting $\bar\lambda(t)$ conforms to $\lambda(t)$ before separatrix crossing can be seen in Figure~\ref{fig:EVOLAM}. There the dotted red trajectory is the unresponsive case with optimal parameters \eqref{eq:RS} and the orange and blue trajectories are for successful and unsuccessful capture respectively. In panel (b), we see that even when zoomed in the evolution of responsive and unresponsive $\lambda$ are indistinguishable up until the time $t_\mathrm{pc}$ when the crossing happens.

\subsection{II.4 Separatrix crossing}
As an adiabatic approximation to $\lambda(t)$, $\bar{\lambda}(t)$ thus provides, for both cases, an estimate for the time at which the system will encounter the separatrix, the ``pseudo-crossing time" $t_\mathrm{pc}$ (so-called in the literature because the exact crossing time depends on post-adiabatic corrections). As derived from adiabatic invariants in Appendix A, the value of $\lambda_\mathrm{pc}=\lambda(t_\mathrm{pc})$, in cases where $\beta^2(\lambda(0))$ is negligible, is given (implicitly, as a function of $p(0)$) by
\begin{equation}\label{pc}
    p(0)=\lambda_\mathrm{pc} + \frac{A_\mathrm{sep}(\lambda_\mathrm{pc})}{4\pi} = \lambda_\mathrm{pc} +\frac{4\beta(\lpc)}{\pi}\;.
\end{equation}
Alternatively, (\ref{pc}) can define the value of $p(0)$ for which the system will encounter the separatrix at a given value of $\lambda=\lambda_\mathrm{pc}$, and thus at a given time $t=t_\mathrm{pc}$. 

As Figure~\ref{outcomes} shows, $t_\mathrm{pc}$ actually is very close, for both responsive and unresponsive cases of $\lambda$, to the time at which the system either enters the separatrix, with $p(t)$ thereafter oscillating around $\lambda$ as in the orange trajectory, or else is bypassed by the moving separatrix, with a one-time downward shift in $p$ followed by ``speed bump'' oscillations around constant $p$. And as Figure~\ref{fig:EVOLAM} shows, $\lambda(t)$ and $\bar\lambda(t)$ remain very close to each other up to and around $t_\mathrm{pc}$. Furthermore, both responsive (Figure~\ref{outcomes}~(a)-(b)) and unresponsive parameters (Figure~\ref{outcomes}~(c)-(d)) do allow the dynamical transition into the separatrix, with success in the dragging control task, while also allowing failure in the control task, depending on precise initial conditions $q(0),p(0)$. Nevertheless the relative proportion of captured trajectories for responsive and unresponsive cases are drastically different, as can be seen in Figure~\ref{fig:ANANUM}.

\subsection{II.5 Systematic effect of responsiveness}
\begin{figure}[htb]
	\centering
	\includegraphics*[width=0.5\textwidth]{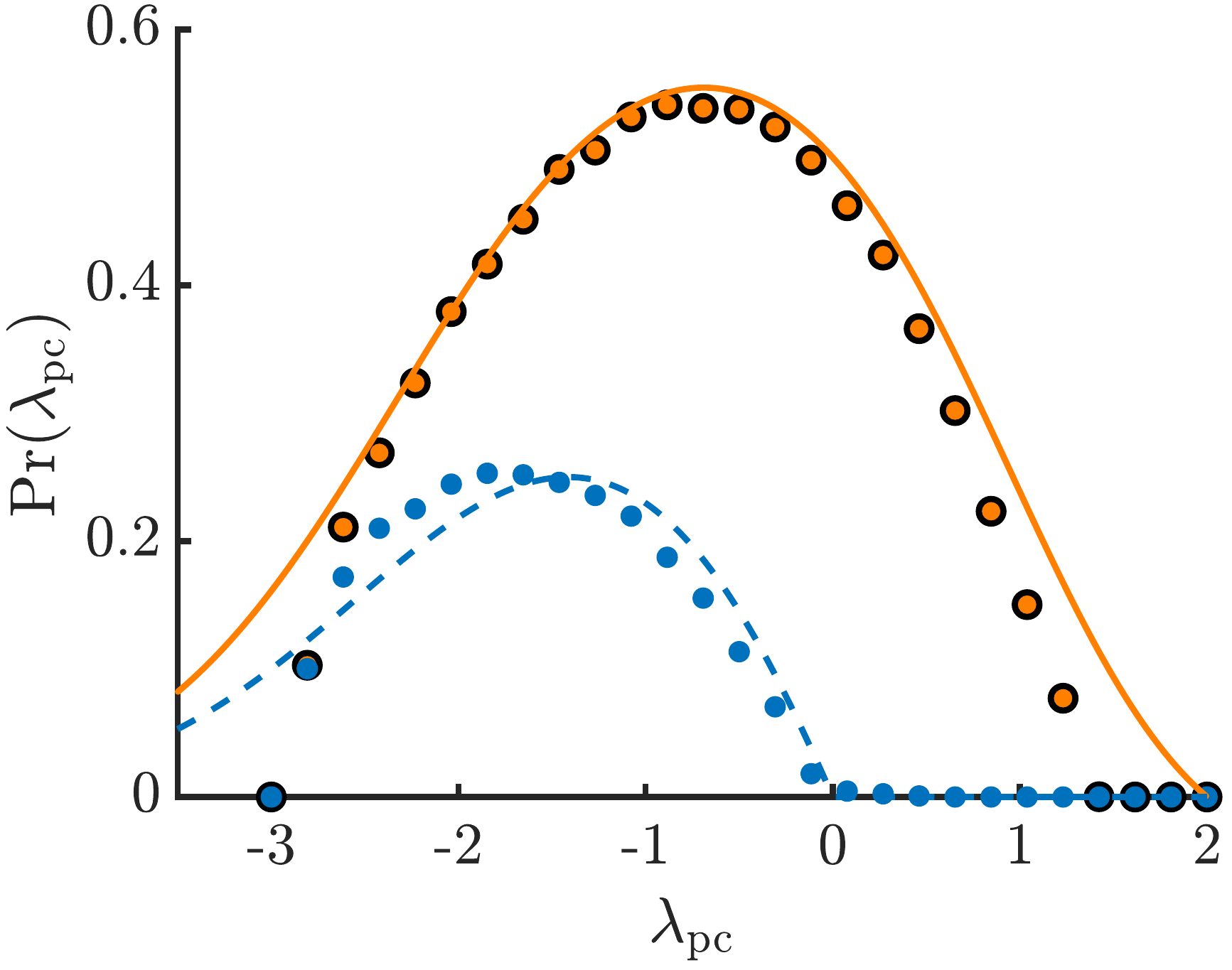}
	\caption{\label{fig:ANANUM} Probabilities of success in the dragging control task, as predicted post-adiabatically by the KNH (blue dashed) and GKNH (orange) formulas, with the parameters from Figure~\ref{outcomes} for different $\lpc$. To calculate each probabilities we evolved $N=10^4$ trajectories numerically for different random initial conditions and determined the probability by counting the trajectories entering the separatrix, $N_\mathrm{sep}$, and then taking the fraction $N_\mathrm{sep}/N$. Blue dots are the probabilities for the unresponsive system while orange dots represent the responsive system.}
\end{figure}

Figure~\ref{fig:ANANUM} represents evolution of an ensemble of $10^4$ initial conditions $q(0)$ randomly distributed in $[-\pi,\pi]$, for each of a range values of $p(0)$ corresponding to 27 different values of $\lambda_\mathrm{pc}$. Trajectories with separatrix crossing are identified for evolution with the unresponsive parameter $\bar\lambda(t)$, and the probability $\mathrm{Pr}(\cdot)$ of control task success is then plotted, as blue dots, versus $\lambda_\mathrm{pc}$. The whole procedure is then repeated with the responsive parameter $\lambda(t)$, and the probabilities shown as orange dots. The results clearly show that the probability of control task success, \textit{i.e.} the fraction of initial conditions leading to orbits with capture into the separatrix, is substantially higher in the responsive case. The probability can even be significant in the responsive case for $\lambda_\mathrm{pc}>0$, where it is zero for the unresponsive case.

\begin{figure*}[htb]
	\centering
	\includegraphics*[width=1\textwidth]{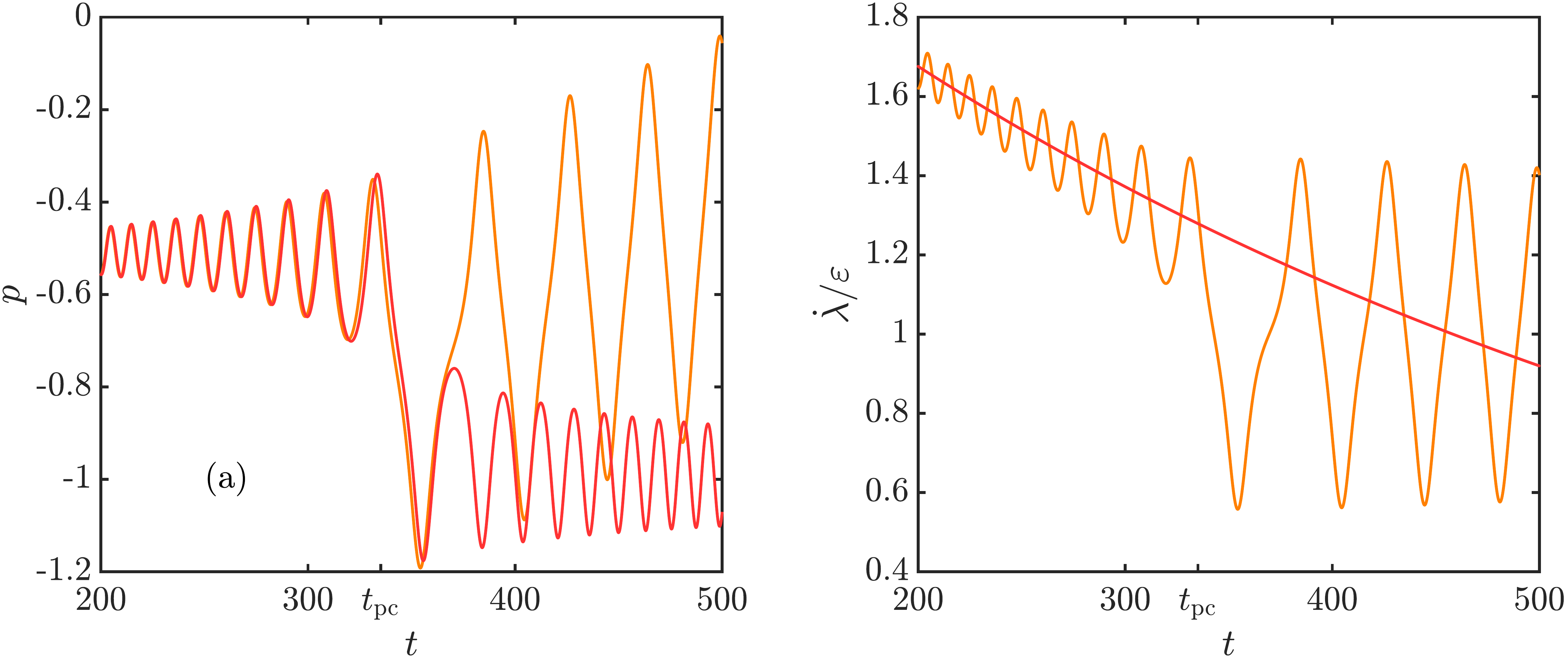}
	\caption{\label{fig:EVODIFF} Evolution for $p$ and $\dot\lambda$ in both the responsive (orange) and unresponsive (red) case for the same initial conditions with the same parameters as before. Note how the $p(t)$ oscillations are very similar for the two cases up until $t_{pc}$, but the oscillations in the responsive orange $\dot{\lambda}(t)$ closely follow the oscillations in $p(t)$ for $t<t_{pc}$, while the unresponsive red curve for $\dot{\lambda}(t)\to \dot{\bar{\lambda}}(t)$ does not. This difference in the responsive $\dot{\lambda}(t)$ behavior is small, but systematically cooperative in triggering capture.}
\end{figure*}

The reason for the great difference in control effectiveness between the responsive and unresponsive cases can be seen in Figure~\ref{fig:EVODIFF}. Because the slight differences between $\lambda(t)$ and $\bar{\lambda}(t)$ are mainly on the short time scale that is normally ignored in adiabatic approximations, in panel (b) of Figure~\ref{fig:EVODIFF} the differences are enhanced by plotting the time derivatives $\dot{\lambda}$ and $\dot{\bar{\lambda}}$. When we can thereby actually see the differences between $\lambda(t)$ and $\bar\lambda(t)$, we notice that the small difference between $\lambda(t)$ and $\bar\lambda(t)$ happens to be that the responsive $\lambda(t)$ has small high-frequency oscillations which closely follow the oscillations of $p(t)$---while the unresponsive $\bar{\lambda}(t)$ does not track these small oscillations of $p(t)$. The difference between responsive $\lambda(t)$ and unresponsive $\bar{\lambda}(t)$ is thus small, but it is by no means simply a random small difference. Instead there is a small \emph{correlation} between $\lambda(t)$ and $p(t)$, which is missing from $\bar{\lambda}(t)$. This small \emph{cooperative} interaction between $p(t)$ and the responsive $\lambda(t)$ can have decisive effects on the unstable evolution around the separatrix.

Since the separatrix includes an unstable fixed point, it should not really be surprising that small differences can affect separatrix crossing dramatically. What is less obvious is that the effect of parameter responsiveness is in fact \textit{systematic}, because $\lambda(t)$ and $p(t)$ interact cooperatively and are therefore correlated. For any single trajectory the success or failure of the separatrix-crossing control task still depends sensitively on precise values of $q(0)$ and $p(0)$, as well as on $\alpha$, $\varepsilon$, and $\lambda(0)$. Over a broad enough ensemble of initial conditions, however, the dynamical responsiveness of $\lambda(t)$ to the system, in comparison to the unresponsiveness of the predetermined $\bar{\lambda}(t)$, contributes a substantial bias in favor of successful capture into the separatrix, as clearly shown in Figure~\ref{fig:ANANUM}.

\subsection{II.6 Generality}
This effect is not just a lucky special feature of this particular model system; neither is it an irreducibly complex manifestation of unstable dynamics which can only be represented accurately by solving equations of motion exactly for all initial conditions. A powerful and general theory provides the solid orange and dashed blue curves in Figure~\ref{fig:ANANUM}, to which the ensemble numerical dots conform closely (except at the edges of the curves, where the adiabatic assumptions of the theory break down for our illustrative parameters). In the next Section we review this theory and show how it implies a generic advantage of open-system control of dynamical transitions.

\section{III. The generalized KNH theorem}
Here we review the Kruskal-Neishtadt-Henrard formula, showing how it applies to our unresponsive case from Section II. We then briefly explain the generalized GKNH version of the result, and show how it predicts the advantages that we have found for the responsive case.
\subsection{III.1 The KNH theorem}
Our unresponsive case is simply a Hamiltonian with a fixed (and slow) time dependence; the probabilities that result from breakdown of adiabaticity around an unstable fixed point are given by the Kruskal-Neishtadt-Henrard theorem \cite{Liouvillepaper}. The KNH theorem is not trivial to prove rigorously \cite{Neishtadt3}, but in essence it follows from Liouville's theorem. If phase space regions are changing their area, the Liouvillian incompressibility of phase space flow along system orbits implies that trajectories must move between regions. For the three areas $A_\downarrow,A_\mathrm{sep},A_\uparrow$ in our pendulum-like system, this leads directly to the simple formula for the probability $\Pr(\cdot)$ of control task success (fraction of initial orbits which cross the separatrix)\begin{equation}
    \Pr(t_\mathrm{pc})=\frac{\dot{A}_\mathrm{sep}(t_\mathrm{pc})}{\dot{A}_\mathrm{sep}(t_\mathrm{pc})+\dot{A}_\downarrow(t_\mathrm{pc})},\label{KNH}
\end{equation}
where the areas $A_{\downarrow,\uparrow}$ are made finite with a constant cutoff far away from the separatrix. In cases where $\Pr(t_\mathrm{pc})$ as given by (\ref{KNH}) is greater than one or negative, the theorem specifies that $\Pr$ is to be taken instead to be one or zero, respectively.

We can use \eqref{KNH} to determine the capture probability for our unresponsive system. The separatrix area \eqref{Asep} leads to \begin{equation}
    \dot{A}_\mathrm{sep}=-8\alpha\dot{\lambda}\lambda \beta.
\end{equation}
The area below the separatrix is $A_\downarrow=2\pi(\lambda-C_\downarrow)-A_\mathrm{sep}/2 $, where $C_\downarrow$ is a constant cutoff far below the separatrix. This yields \begin{equation}
    \dot{A}_\downarrow=2\pi\dot{\lambda}-4\alpha\dot{\lambda}\lambda\beta.
\end{equation}

The probability of entering the separatrix in the unresponsive case is then given by the formula \begin{equation}
    \Pr(t_{pc})=\frac{-4\alpha\lambda\beta}{\pi-2\alpha\lambda\beta} \bigg|_{\lambda=\lpc}\;.
\end{equation}  It is noteworthy that this expression is independent of $\dot{\lambda}$, being obtained in the adiabatic limit; we could have chosen $\bar\lambda(t)$ arbitrarily (as long as it is sufficiently slow) and obtained the same capture probability. This formula corresponds to the dashed blue line in Figure~\ref{fig:ANANUM}, which agrees well with the blue dots from the numerical calculation. It also explains why there is no capture for $\lpc>0$, where the separatrix area is constantly shrinking and no states can enter the separatrix due to Liouville's theorem.
The only way to exceed this KNH capture efficiency is to evade Liouville's theorem, by having an open system with some non-Hamiltonian term in its evolution.

\subsection{III.2 The generalized KNH theorem}
The KNH theorem has been generalized to include some forms of non-Hamiltonian evolution. In this section we will adapt the formulas from \cite{Neishtadt2,Neishtadt3} for the probability of entering a double-well-like system with Hamiltonian $H(\boldsymbol{\lambda},q,p)$, tailoring them to our example, and then give a heuristic derivation of the formula. In \cite{Neishtadt3} it is explicitly anticipated that such adaption to different phase space geometries is possible, even to those on a cylinder as in our case.

Firstly we note that our example fits the general form of equation of motion\begin{align}
    \dot{\lambda}=&\varepsilon f_\lambda(\lambda,q,p,),\label{generalLAM}\\
    \dot{q}=&\pfrac{H}{p}+\varepsilon f_q(\lambda,q,p),\label{generalQ}\\
    \dot{p}=&-\pfrac{H}{q}+\varepsilon f_p(\lambda,q,p),\label{generalP}
\end{align} for which the theorem holds.
We further have a phase space structure like that of a physical pendulum: an eye-like inner region bounded by a separatrix consisting of upper and lower arcs. The separatrix is a contour of constant instantaneous energy; for convenience (and without loss of generality) we set it to have zero energy at all times, so that the energy in its interior is negative. Any trajectory which crosses the separatrix will first approach it and orbit past it closely for some time; we will compute and compare the loss of energy $h$ during orbits just above and just below the separatrix, using the upper-arc ``frown'' and lower-arc ``smile''  symbols as subscripts to denote orbits above and below the separatrix, respectively. 

\begin{figure}[htb]
	\centering
	\includegraphics*[trim={10 50 0 50},clip,width=0.5\textwidth]{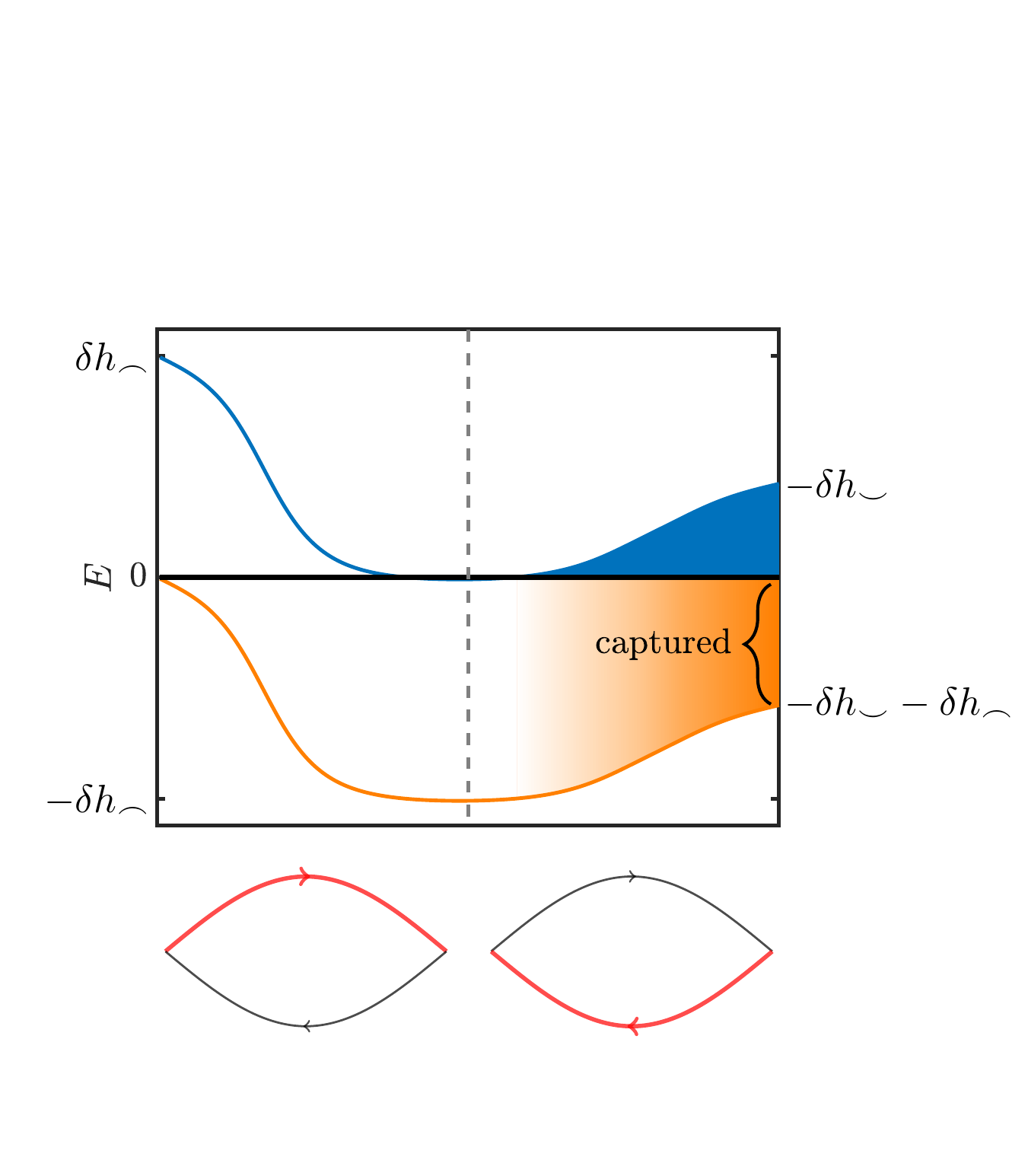}
	\caption{\label{fig:EnergyEvo} Change of the energy interval $[0, \delta h_\frown]$ (upper bound blue, lower bound orange) over time, during the evolution along upper and lower arc of the separatrix. Along the upper arc the energy $\delta h_\frown$ gets lost, yielding interval $[-\delta h_\frown, 0]$. Along the lower arc, the energy $-\delta h_\smile$ is gained; orbits which thereby rise above zero energy again are not returning to their initial phase space region above the separatrix, however, but are rather proceeding below the lower arc of the separatrix, where energy is also positive. These blue-shaded orbits will therefore \emph{not} cycle around the separatrix again and be captured; instead they have successfully crossed the whole negative energy region, climbing back up to positive energy, and will never be captured. The captured fraction is what remains below the separatrix energy, colored in orange.}
\end{figure}

In particular we consider a set of orbits which are initially above the separatrix with energies in the range $(0,\delta h_\frown]$; see Figure~\ref{fig:EnergyEvo}. This set (except an exponentially small subset near the fixed point \cite{Henrard}) will evolve closely to the separatrix and will pass along it much faster than the separatrix will deform. The upper bound $\delta h_\frown$ is the energy loss of the trajectories, calculated as integral over the upper arc of the instantaneous separatrix contour. This energy loss over the upper separatrix arc will bring the upper energy of our orbits' range down to the separatrix energy of zero. This means that we are dealing with the largest set of orbits which begins above the upper branch of the separatrix and is brought entirely under it, by the parameter change and non-Hamiltonian energy loss, as it follows close to the arc. The energy range $\delta h_\frown$ can be computed in the adiabatic limit by integrating along the upper arc:
\begin{equation}
    \delta h_\frown=-\intfrown\d{\tau} \dfrac{H}{\tau}=-\varepsilon\intfrown\d{\tau} \Big(\pfrac{H}{\lambda}f_\lambda+\pfrac{H}{q}f_q+\pfrac{H}{p}f_p\Big)\label{dHfrown}\end{equation}
where $\tau$ refers to time evolution under the instantaneous Hamiltonian at $t_\mathrm{pc}$ with ``frozen" $\lambda=\lpc$. 

This means that all states above the separatrix with energies in the interval $(0,\delta h_\frown]$ will transition to $(-\delta h_\frown,0]$, as sketched in the first half of Figure~\ref{fig:EnergyEvo}. To determine to which region of phase space the state will transition, we follow its path close to the lower arc of the separatrix.
There the energy changes by
\begin{equation}
    \delta h_\smile=-\intsmile\d{\tau} \dfrac{H}{\tau}=-\varepsilon\intsmile\d{\tau} \Big(\pfrac{H}{\lambda}f_\lambda+\pfrac{H}{q}f_q+\pfrac{H}{p}f_p\Big).\label{dHsmile}
\end{equation} This leads to a spread in energy of $(-\delta h_\frown-\delta h_\smile,-\delta h_\smile]$, as sketched in the second half of Figure.~\ref{fig:EnergyEvo}. It is essential to realize that the blue-shaded region in Figure~\ref{fig:EnergyEvo} has risen in energy above zero, so that it has escaped from inside the separatrix, but it has \emph{not} returned to its starting region above the separatrix! Instead it has fallen below the lower arc of the separatrix, where the energy again rises. This blue region therefore represents a portion of the ensemble which has crossed through the negative-energy region without being trapped in it: it is the portion of the initial energy range just above the separatrix which does \emph{not} get captured. 

If for some values of $\lpc$ there is no blue region, because none of the initial energy range ever comes back to positive energies, then capture is certain; if there is no orange region, capture is impossible. Otherwise, the fraction of orbits near the separatrix which are captured inside it is given by the ratio of the final energy width of the orange-shaded region in Figure~\ref{fig:EnergyEvo} to the initial energy width $\delta h_\frown$, 
\begin{equation}
    \Pr(\lpc)=\frac{\delta h_\frown+\delta h_\smile}{\delta h_\frown}\bigg|_{\lambda=\lpc}.   \label{eq:GKNH}
\end{equation}

To see how these generalizations of the KNH probability formula apply to our responsive case, we rewrite the integrals \eqref{dHfrown} and \eqref{dHsmile} and decompose $f_\lambda$. We change the integration variable from $\tau$ to $q$  and use $f_q=f_p=0$, obtaining
\begin{equation}
     \delta h_\frown=-\varepsilon\int\limits_{-\pi}^{\pi}\d{q} \pfrac{H}{\lambda}\left(\pfrac{H}{p}\right)^{-1}f_\lambda(\lambda,q,p_+(q)),
 \end{equation}
 with $p_+$ from \eqref{eq:psep}. The two partial derivatives in the integral evaluate to \begin{equation}
\pfrac{H}{\lambda}\left(\pfrac{H}{p}\right)^{-1}=-1\mp2\beta^\prime\cos(q/2) \label{prefacor}
\end{equation} for the two different paths, where $\beta^\prime=\partial_\lambda\beta$.
 We can then use a decomposition to separate the effects\begin{equation}
     f_\lambda(\lambda,q,p_\pm(q))=\bar{f}(\lambda)+\Delta_\pm(\lambda,q).
 \end{equation}
 The function $\bar{f}$ is simply a time dependence of the Hamiltonian, so in the adiabatic limit it has a constant value at $\lambda=\lpc$, and we can pull it in front of the integrals for both $\delta h_\frown$ and $\delta h_\smile$. In the numerator of the probability formula this then leaves us with \begin{align}
     \varepsilon\bar{f}\int\limits_{-\pi}^\pi \d{q} 1+2\beta^\prime\cos\left(\frac{q}{2}\right)-1+2\beta^\prime\cos\left(\frac{q}{2}\right)&=16\varepsilon\bar{f}\beta^\prime\nonumber\\
     =\varepsilon\bar{f}\pfrac{}{\lambda} A_\mathrm{sep}.
 \end{align}
 If $\Delta=0$ we have  $\varepsilon\bar{f}\partial_\lambda=\dot{\lambda}\partial_\lambda=\partial_t$ and the rates of change of the areas from the KNH theorem are recovered.
 
 To evaluate the term arising from $\Delta_\pm$ in the numerator, we choose the decomposition $\Delta_\pm(\lambda,q)=\gamma(p_\pm(q)-\lambda)$. The term arising from the back-reaction is then just 
 \begin{equation}
     \varepsilon\int\limits_{-\pi}^\pi\d{q} (\Delta_+-\Delta_-)=16\varepsilon\gamma\beta=\varepsilon\gamma A_\mathrm{sep}, \label{newterm}
 \end{equation} where $\Delta_-$ is subtracted because the path of integration reverses and the second summand in from \eqref{prefacor} vanishes because of symmetry. From \eqref{newterm} we immediately see the mechanism of our back-reaction by which the probability is increased: terms proportional to the separatrix area are added to the numerator, allowing it to be greater than zero even for a shrinking separatrix.

A more detailed calculation of the values of $\delta h_\frown,\delta h_\smile$ can be found in Appendix C. The final formula for the probability is \begin{equation}
    \Pr(\lpc)=\frac{\gamma A_\mathrm{sep}+A_\mathrm{sep}^\prime}{A_\downarrow^\prime +A_\mathrm{sep}^\prime +\gamma A_\mathrm{sep}/2-4\pi\gamma A_\mathrm{sep}A_\mathrm{sep}^\prime/16^2}\bigg|_{\lambda=\lpc},
\end{equation} or explicitly in terms of the system parameters for our model
 \begin{equation}
    \Pr(\lpc)=\frac{16\gamma\beta-8\alpha\lambda\beta}{2\pi +8\gamma\beta+4\alpha\lambda\beta-2\pi\gamma\alpha\lambda\beta^2}\bigg|_{\lambda=\lpc}.
\end{equation}
This provides the theoretical curve for the responsive case (orange) in Figure~\ref{fig:ANANUM}.

In Figure~\ref{fig:ANANUM} we also see some deviations of the analytical probabilities at the fringes of the distribution. These arise when the change of the separatrix area is slow compared to the speed of its instantaneous fixed point, so that the adiabatic limit no longer holds well.

\subsection{III.3 Optimal capture}
We have seen in the last section that what matters for capture is the behaviour of the $\mathbf{f}=(\FL,\FQ,\FP)^\top$ terms along the separatrix. Now we will assume that we have at least some control over $\mathbf{f}$, and ask whether and how a careful choice of $\mathbf{f}$ can make control success certain, \emph{i.e.} raise $\Pr$ to one. Alternatively, it could be that capture inside the separatrix is a hazard that we would like to avoid; in such a case we would try to bring $\Pr$ to zero.


Let us consider that the separatrix at the pseudocrossing parameter value $\lpc$ has been given. If we can choose $\mathbf{f}$ at every point along the separatrix, how should we chose it to enhance capture, or suppress it? A glance at \eqref{eq:GKNH} reveals that we want to minimize $\delta h_\smile$ and maximize $\delta h_\frown$---or else the opposite, if we desire to avoid having the system get captured.

In either case we can rewrite \eqref{dHfrown} as \begin{equation}
    \delta h_\frown=-\intfrown\d{\tau} \mathbf{n}\cdot \mathbf{f},
\end{equation} where $\mathbf{n}(q(\tau),p(\tau))=(\partial_\lambda H,\partial_q H,\partial_p H)^\top$ is the outward-pointing normal vector of the surface that the separatrix sweeps out as $\lambda$ is varied, and the dot is the usual scalar product. An analogous formula holds for $\delta h_\smile$. Hence the goal of increasing $\Pr$ means increasing $\mathbf{n}\cdot\mathbf{f}$ over the upper branch of the  separatrix, and decreasing it over the lower branch. If instead the goal were to decrease $\Pr$, because capture inside the separatrix was a danger rather than a desired outcome, then we would want to decrease $\mathbf{n}\cdot\mathbf{f}$ over the upper branch of the separatrix and decrease it over the lower.


Our ability to control $\mathbf{n}\cdot\mathbf{f}$ at any point in phase space may in practice be limited. In the example from the previous subsection, only $f_\lambda$ was non-zero; even if our control over $\mathbf{f}$ were limited by this constraint, we could in principle still make $\FL$ positive on the upper arc and negative on the lower arc, so that $\delta h_\smile>0$ while $\delta h_\frown<0$, which then immediately makes the right-hand side of (\ref{eq:GKNH}) greater than one, meaning $\Pr = 1$. Or else we could make $\FL$ sufficiently negative on the lower arc, and sufficiently positive on the upper arc, to make $\delta h_\smile<0$ while $\delta h_\frown + \delta h_\smile >0$, so that the right-hand side of (\ref{eq:GKNH}) becomes negative, meaning $\Pr=0$. 

A responsive parameter can thus in principle achieve perfect control, either to guarantee that the system will be captured inside the separatrix, or to guarantee that it will not be. Whether those ideals can be achieved in practice depends on how well $\mathbf{f}$ can in fact be engineered. 



\section{IV. Discussion}
As our example in Section II illustrates, and our derivation in Section III from the GKNH shows in general, the efficiency of an adiabatic sweep type of control procedure can be altered by giving the control parameter some dynamical responsiveness, so that it suffers back-reaction from the target system instead of exactly following a pre-programmed sweep protocol. In our example in Section II we showed dramatic increase, in the responsive case, in the probability of capturing the system into a target region of phase space. Our derivation in Section III indicates, however, that capture probabilities can in general be either raised or lowered, depending on exactly how the control parameter responds to the system. 

In some cases where control parameter back-reaction is lowering the capture probability, this may be an unwanted effect which needs to be eliminated; our message in such cases is that even small back-reaction from the system onto the control parameter could be the cause of otherwise mysteriously low capture efficiency. In other cases, however, capture might not be a goal but an obstacle; the control task might be to avoid capture of the system into an undesired state. Here our message is that an appropriate form of parameter back-reaction can indeed reduce the chances of undesired capture. Conversely, desired capture can have its probability enhanced, potential up to unity, by appropriately engineered responsive sweeping.

In either case we emphasize that the effect of parameter responsiveness on capture probabilities can be substantial even when the parameter back-reaction itself is a small effect, because capture by separatrix crossing is a sensitive process. The subtlety of this effect has indeed sometimes been overlooked, for example in celestial mechanics analysis of satellite capture \cite{Goldreich, Henrard}. Problems of this nature have been studied in celestial mechanics for decades, and yet the generalized Kruskal-Neishtadt-Henrard theorem has only quite recently been given a rigorous proof \cite{Neishtadt3}. The main purpose of our paper has been to emphasize the implications of these results for more general systems than celestial ones, including laboratory set-ups in atomic and optical physics, where various forms of open-system dynamics can indeed be engineered.

Even microscopic systems may require efficient control mechanisms, whether artificially designed or naturally evolved. Adiabatic control procedures can be robust and efficient; a major advantage is that they can achieve blind control, in the sense that adequate success probabilities are achieved without precise monitoring or control of fast degrees of freedom \cite{Liouvillepaper}. Adding a bit of responsiveness to a control parameter can substantially improve this kind of blind control, inasmuch as the parameter itself may be less blind the system's fine-grained behavior than the experimenter is.

Taken to the extreme this idea becomes obvious. A Maxwell's Demon is a hypothetical controller which affects the system in drastically different ways depending on the state of the system; at least if one is careless enough in formulating the Demon, it can seem even to break the Second Law of Thermodynamics. Intelligent manipulation of a system, to the point where one carries it by hand to its target state, can obviously succeed at control tasks quite well. 

Our point here, however, is that even very simple and limited control responsiveness, far short of intelligence, can guide system evolution efficiently. Leakage of system information into the environment is often seen as an enemy which must be fought by isolating the system more perfectly; we argue here to the contrary, that open system control, in which the system is allowed to affect its environment, can offer dramatic benefits. Dissipation is usually considered an unwanted energy loss, but cooling an engine is crucial to its performance---and this principle can extend even to microscopic active particles \cite{DissDaemon}.

The particular form of parameter responsiveness which we analysed in Section II was not chosen as a finely tuned example to show performance far better than can realistically be achieved without microscopic design. On the contrary it is an example of mindlessly simple open system control: inserting the canonical equation of motion obtained by \eqref{eq:HAM} together with \eqref{eq:lamdotExact} into one another leads immediately to
\begin{equation}
    \ddot{q}=-\beta^2\sin q -\varepsilon(1+\gamma\dot{q})\;,
\end{equation}
the equation of motion of a particle in a tilted lattice potential with damping. Even the crudest form of open system control can work surprisingly well.

\section{Acknowledgments}
The authors acknowledge support from State Research Center OPTIMAS and the Deutsche Forschungsgemeinschaft (DFG) through SFB/TR185 (OSCAR), Project No. 277625399.

\bibliographystyle{unsrt}
\bibliography{mainbib} 

\section*{Appendix}
\appendix

\section{A. Averaging $\lambda$}\label{App: B}
The unresponsive system is the time-averaged system obtained by the averaging principle \cite{Neishtadt3,Freidlin2}, where we denote quantities with a bar. Since the initial ensemble is above the separatrix starting at $p(0)$ and an open $2\pi$-periodic orbit, we first obtain the averaged momentum \begin{equation}
    \dot{\bar{p}}=\frac{-1}{T}\int\limits_0^T\d{t} g\exp\left(-\frac{\alpha}{2}\lambda^2\right)\sin(q)=0,
\end{equation} so that $\bar{p}(t)=p(0)$. To average $\lambda$, we focus on open orbits before capture. We integrate \eqref{eq:lamdotExact} over one period $T$, which yields \begin{equation}
    \dot{\bar{\lambda}}=\frac{\varepsilon}{T}\int\limits_0^T\d{t} 1+\gamma(p-\lambda)=\varepsilon(1+\gamma(\bar{p}-\bar{\lambda})).\label{eq:DerivativeAvUp}
\end{equation}
Since $\bar{p}=p(0)=\mathrm{const}$ we can integrate this and get the solution
\begin{equation}
    \bar{\lambda}=\left(\lambda(0)-\frac{1+\gamma p(0)}{\gamma}\right)\mathrm{e}^{-\varepsilon\gamma t}+\frac{1+\gamma p(0)}{\gamma}. \label{eq:TimeAvUp}
\end{equation}

\section{B. Calculating $t_\mathrm{pc}$ from the Averaging Principle}\label{App: A}
Here we calculate the slow evolution of one of the first integrals of \eqref{eq:HAM}. We define the action in the frozen system to be \begin{equation}
    I=\oint\d{q} p, 
\end{equation}where we either integrate from $0$ to $2\pi$, since we are only interested in times before the separatrix crossing occurs. From \begin{equation}
    \dot{\bar{p}}=\frac{-1}{T}\int\limits_0^T\d{t} g\exp\left(\frac{\alpha}{2}\lambda^2\right)\sin(q)=0,
\end{equation} we can directly see that \begin{equation}
    I=2\pi p(0)=\mathrm{const}.
\end{equation}
The orbit can only be outside the separatrix as long as the area under the upper arc of the separatrix, $A_\frown$ is lower than the initial action. The area below the separatrix is \begin{equation}
    A_\frown(\lambda)=2\pi\lambda+A_\mathrm{sep}(\lambda)/2.
\end{equation}
The pseudocrossing parameter value is where the condition \begin{equation}
    2\pi p(0)=A_\frown(\lpc).
\end{equation} Having $\lpc$ the inversion of \eqref{eq:TimeAvUp} yields the pseudocrossing time $t_\mathrm{pc}$. Or for the numerical calculations, we can choose the initial condition \begin{equation}
    p(0)=\lpc+4\beta(\lpc)/\pi
\end{equation} to get a trajectory with pseudocossing parameter value $\lpc$.

\section{C. Calculating the Rate from $A_\downarrow$ for the Responsive System}\label{App: C}
Here we calculate an expressions for $\delta h_\frown$ and $\delta h_\smile$ from \eqref{dHfrown} and \eqref{dHsmile}. In our case with vanishing $f_q,f_p$
\begin{equation}
    \delta h_\frown=-\varepsilon\intfrown\d{t} \pfrac{H}{\lambda}f_\lambda=-\varepsilon\int\limits_{-\pi}^{\pi}\d{q} \pfrac{H}{\lambda}f_\lambda/\dot{q}.
\end{equation}
We have \begin{align}
    \dot{q}&=p-\lambda,\\
    p(q)&=\lambda+ \beta\sqrt{2(1+\cos q)}=:\lambda+ \beta S\\
    \pfrac{H}{\lambda}&=\lambda-p-2\beta\beta^\prime(\cos(q)+1),\\
    &=\lambda-p-\beta^\prime (p-\lambda)^2/\beta\\
    f_\lambda&=1+\gamma(p-\lambda),
\end{align} where $\beta^\prime$ is the derivative of $\beta$ with respecto to $\lambda$.
Inserting this yields 
\begin{align}
    \delta h_\frown&=-\varepsilon\int\limits_{-\pi}^{\pi}\d{q} (\lambda-p-\beta^\prime (p-\lambda)^2/\beta)\frac{1+\gamma(p-\lambda)}{p-\lambda},\\
    &=\varepsilon\int\limits_{-\pi}^{\pi}\d{q} (1+\beta^\prime S)(1+\gamma\beta S).
\end{align}
With \begin{align}
    &\int\limits_{-\pi}^{\pi}\d{q} S=8,\\
    &\int\limits_{-\pi}^{\pi}\d{q} S^2=4\pi,\\
    &\beta^\prime=-\frac{\alpha\lambda}{2}\beta,
\end{align}
follows \begin{equation}
    \delta h_\frown/\varepsilon=2\pi+8\gamma\beta-4\alpha\lambda\beta-2\pi\gamma\alpha\lambda\beta^2.
\end{equation}

By letting go $S\to-S$ and gaining a global minus sign by reversing the path of integration we get\begin{equation}
    \delta h_\smile/\varepsilon=-2\pi+8\gamma\beta-4\alpha\lambda\beta+2\pi\gamma\alpha\lambda\beta^2.
\end{equation}

\end{document}